# Steady and dynamic magnetic phase transitions in interacting quantum dots arrays coupled with leads

Hang Xie*, Xiao Cheng, Xiaolong Lü

Department of Physics, Chongqing University, Chongqing, China

*Email: xiehangphy@cqu.edu.cn

We apply the Hubbard model, non-equilibrium Green's function (NEGF) theory, exact diagonalization (ED) and the hierarchical equations of motion (HEOM) method to investigate abundant magnetic phase transitions in the 1D interacting quantum dots arrays (QDA) sandwiched by non-interaction leads. The spin polarization phase transitions are firstly studied with a mean-field approximation. The many-body calculation of the ED method is then used to verify such transitions. We find with the weak device-leading couplings, the anti-ferromagnetic (AF) state only exists in the uniform odd-numbered QDA or the staggered-hopping QDA systems. With increasing the coupling strength or the bias potentials, there exists the magnetism-to non-magnetism phase transition. With the spin-resolved HEOM method we also investigate the detailed dynamic phase transition process of these lead-QDA-lead systems.

## I. INTRODUCTION

The magnetism in nano structures attracts a lot of research interests in recent years [1]. For example, in the zigzag edges of graphene nanoribbons, there exists the spontaneous spin polarization with anti-ferromagnetic (AF) order [2, 3]. Such spin polarization also appears in the graphene nano-patch, nano-hole or other non-uniform structures [4-6]. With some external magnetic fields, these spin polarization may be in a ferromagnetic (FM) order, or even in a periodic oscillation as some spin density waves (SDW) on graphene nanoribbons [3, 7].

Besides the graphene nanoribbons, the magnetic order also exists in the one-dimensional structures, such as carbon atom chain (CAC), or carbene and polyacetylene system. There are many researches on CAC based on the first-principle calculations [8, 9]. Due to the Peierls distortion, CAC may have the cumulene structure with the same double bonds (=C=C=) or polyyne structure with the alternative single and triple bonds (-CΞC-) in different conditions. S. Cahangirov and Z. Zanolli's works also show that the even-number CAC often has zero or very small magnetic moment while the odd-number CAC has a much large magnetic moment (with the AF order). In the polymer systems, S.J. Xie's work shows the colossal magnetoresistance in the magnetic-lead-polymer structures[10].

These magnetic phenomena can be explained by the Hubbard model. Hubbard model describes the Coulomb interaction by the on-site repulsion term. A self-consistent calculation with a mean-field (MF) approximation can easily give the AF order for the odd-numbered tight-binding (TB) chain. The AF order is reasonable for this Hubbard model. It is because that for the odd number of sites, there has to be different spin-up and spin-down orbitals below the Fermi level. So the spin polarization is an inevitable outcome for the ground state, which also obeys the Lieb's theorem [11].

Hubbard model is a very important model for the strong interacting systems. Based on the Bethe ansatz, the analytical solution of 1D infinite Hubbard chain was obtained by Wu's work[12]. There are abundant physical phenomena in the 1D Hubbard model, such as the Mott transition, Coulomb blockade and Kondo resonance [13, 14].

Most of these Hubbard model works focus on the isolate structure or the periodic systems. The open system (the interacting structure with two semi-infinite leads) includes infinite number of particles, which can be handled by the non-equilibrium Green's function (NEGF) theory with the MF approximation [15, 16]. It is worthy and interesting to study the magnetic properties for this open system with 1D interacting structures. To avoid the Peierls distortion of CAC, we use the QDA in our study. We find there exist a new type of magnetic phase transition in the open system. To best of our knowledge, there has no such report for this type of phenomena.

Besides the mean-field approximation, exact diagonalization is also an important tool in the Hubbard model study. For instance, G.L. Chen et al, and B. Muralidharan et al used this method, combined with the rate equation to study the resonant tunneling through quantum dots arrays (QDA)[13, 17]. Nguyen H. Le et al recently use the extended Hubbard model to investigate the mesoscopic transport of QDA system [18]. H. Ishida et al used ED and dynamical mean field theory (DMFT) to study the electron structures of the Hubbard molecule with leads[14]. In this paper we also use ED to test these magnetic phenomena.

To give a deep understanding of this magnetic phase transition, we also use the time-dependent quantum transport calculation: the hierarchical equation of motion (HEOM) method for this system [19-21]. In this work we generalized our previous HEOM theory to the spin resolved case. This calculation gives a dynamic description for the transition process. We also find some large bias potential can also induce such magnetic-to-nonmagnetic phase transition.

## II. THEORIES AND MODELS

We use the spin-resolved TB model to investigate the magnetic phase transition in the 1D QDA in this paper. The Hubbard model is used for the Coulomb interaction, which has the expression below

$$H = \sum_{<i,j>} t_{i,j} c_{i\uparrow}^{\dagger} c_{j\uparrow} + \sum_{<i,j>} t_{i,j} c_{i\downarrow}^{\dagger} c_{j\downarrow} + U \sum_{i} n_{i\uparrow} n_{i\downarrow} \quad (1)$$

where $t_{i,j}$ is the hopping integral between atom site i and j, $c_{i\uparrow}^{\dagger}$ and $c_{i\uparrow}$ are the creation and annihilation operator of spin-up electron; $c_{i\downarrow}^{\dagger}$ and $c_{i\downarrow}$ are for the spin-down case. U is the Hubbard constant, which means the on-site Coulomb repulsion energy between the spin-up and spin-down electrons. $<i,j>$ means the summation is calculated among the nearest neighboring sites.

## A. NEGF theory with Hubbard model

Here we use the mean-field approximation in the Hubbard model for a single-electron Hamiltonian.

$$H = \sum_{<i,j>} t_{i,j} c_{i\uparrow}^{\dagger} c_{j\uparrow} + \sum_{<i,j>} t_{i,j} c_{i\downarrow}^{\dagger} c_{j\downarrow} + U \sum_{i} [n_{i\uparrow} < n_{i\downarrow} > + n_{i\downarrow} < n_{i\uparrow} > - \frac{1}{2}] \quad (2)$$

where $< n_{i\uparrow} >$ and $< n_{i\downarrow} >$ are the mean-field average electron density of spin-up and spin-down case.

For the open system, the NEGF theory is used for the electron density calculations and transmission spectrum. The Green's function is obtained by the following formula [16]

$$(E\mathbf{I} - \mathbf{H} - \mathbf{\Sigma}^r) \cdot \mathbf{G}^r = \mathbf{I} \quad (3)$$

where $\mathbf{\Sigma}^r$ is the self-energy and I is the unit matrix with the dimension of the device. $\mathbf{\Sigma}^r$ is often calculated by the surface Green's function from the iteration method [16]. In the case of the bias potential, the electron density is obtained from the diagonal part of the density matrix $\rho_i = \sigma_{i,i}$. And $\boldsymbol{\sigma}$ is related to the less Green's function $\mathbf{G}^<$.

$$\boldsymbol{\sigma} = -i\mathbf{G}^<(t=0) \quad (4)$$

From the NEGF theory, $\mathbf{G}^<$ can be calculated by the following formula

$$\mathbf{G}^<(E) = \mathbf{G}^a(E)\mathbf{\Sigma}^<(E)\mathbf{G}^r(E) \quad (5)$$

where the less self-energy $\mathbf{\Sigma}^<(E)$ is related to the line-width function of lead L and R.

$$\mathbf{\Sigma}^< = \mathbf{\Sigma}_L^< + \mathbf{\Sigma}_R^< = i(f_L \mathbf{\Gamma}_L + f_R \mathbf{\Gamma}_R) \quad (6)$$

We may use the residue theory to obtain the integral calculation for the electron density: $\boldsymbol{\sigma} = \int \mathbf{G}^<(E)dE$ [22]. The Padé spectrum decomposition will be utilized to expand the Fermi-Dirac function as the following form [19]

$$f_\alpha(z) = \frac{1}{1+\exp(z)} \approx \frac{1}{2} + \sum_{p=1}^{N_p} \left(\frac{R_p}{z - z_p^+} + \frac{R_p}{z - z_p^-}\right). \quad (7)$$

where $R_p$ and $z_p^+$ are the p[th] residue and Pade pole.

## B. Exact diagonalization

Exact diagonalization or ED is a useful tool to investigate the many-body property of nano systems. In this method, instead of using the common single-electron wavefunction $\psi(\mathbf{r})$, the many-body wavefunction $\Psi(\mathbf{r}_1,\mathbf{r}_2,\cdots)$ and the corresponding many-body eigen problem is considered [18, 23].

In the calculation of ED, the many-body wavefunction of electron is expressed in the occupation number representation. The basis is written as $\Phi_k^{N_\alpha,N_\beta} = |n_1^\alpha, n_2^\alpha \cdots n_{N_\alpha}^\alpha; n_1^\beta, n_2^\beta \cdots n_{N_\beta}^\beta >$, where $n_i^{\alpha(\beta)} = 1 \text{ or } 0$, $N_\alpha$ and $N_\beta$ are the total number of spin-up and spin-down electrons, *k* is the index of the basis. Then the wavefunction is the linear combination of these basis: $\Psi_n = \sum_k a_k^{(n)} \Phi_k^{N_\alpha,N_\beta}$ ((n) is used for the index of the eigen-energies). With the particle number operator: $\hat{n}_i^{\alpha(\beta)} = c_i^{\alpha(\beta)+} c_i^{\alpha(\beta)}$, we may obtain the electron density $n_i^{(n)} = \sum_{i\in\{k\}} |a_k^{(n)}|^2$, where $i \in \{k\}$ means the site of *i* is included in the many-particle state $\Phi_k = |1_1^\uparrow 0_2^\uparrow \cdots 1_i^\uparrow \cdots >$.

## C. HEOM theory

HEOM theory is developed in our previous works [19, 20]. It is suitable for the open quantum systems.

In this work we extend the original HEOM theory to the spin case. The electron density with spin *s* (*s* is for spin up or spin down) is denoted as $\boldsymbol{\sigma}_D^s(t)$. And the corresponding auxiliary density matrices are changes as the spin-resolved ones with the superscript *s*. We here write down the central equations of spin-resolved HEOM

$$i\partial_t \boldsymbol{\sigma}_D^s(t) = [\mathbf{H}_D^s(t), \boldsymbol{\sigma}_D^s(t)] - \sum_{\alpha}^{N_\alpha} \sum_{k=1}^{N_k} (\boldsymbol{\varphi}_{\alpha k}^s(t) - \boldsymbol{\varphi}_{\alpha k}^{s\dagger}(t)) \tag{8}$$

$$i\partial_t \boldsymbol{\varphi}_{\alpha k}^s(t) = [\mathbf{H}_D^s(t) - i\gamma_{\alpha k}^+ - \boldsymbol{\Delta}_\alpha(t)]\boldsymbol{\varphi}_{\alpha k}^s(t) - i[\boldsymbol{\sigma}_D^s(t)\mathbf{A}_{\alpha k}^{>+} + \bar{\boldsymbol{\sigma}}_D^s(t)\mathbf{A}_{\alpha k}^{<+}] + \sum_{\alpha'}^{N_\alpha} \sum_{k'=1}^{N_k} \boldsymbol{\varphi}_{\alpha k,\alpha' k'}^s(t) \tag{9}$$

$$i\partial_t \boldsymbol{\varphi}_{\alpha k,\alpha' k'}^s(t) = -[i\gamma^+{}_{\alpha k} + \boldsymbol{\Delta}_\alpha(t) + i\gamma^-{}_{\alpha' k'} - \boldsymbol{\Delta}_{\alpha'}(t)] \cdot \boldsymbol{\varphi}_{\alpha k,\alpha' k'}^s(t)$$

$$+i(\mathbf{A}^{>-}_{\alpha' k'} - \mathbf{A}^{<-}_{\alpha' k'})\boldsymbol{\varphi}^{s}_{\alpha k}(t) - i\boldsymbol{\varphi}^{s\dagger}_{\alpha' k'}(t)(\mathbf{A}^{>+}_{\alpha k} - \mathbf{A}^{<+}_{\alpha k}) . \quad (10)$$

where $\bar{\boldsymbol{\sigma}}^{s}_{D}(t) = \mathbf{I} - \boldsymbol{\sigma}^{s}_{D}(t)$ and $N_k = N_d + N_p$ is the total number of the Lorentzian and Padé poles. $\mathbf{H}^{s}_{D}$ is Hamiltonian of the device with the mean-field approximations for the spin-up or spin-down cases. They are expressed as

$$\mathbf{H}^{\uparrow}_{D} = \sum_{i} V_i(t) c^{\dagger}_{i\uparrow} c_{i\uparrow} + \sum_{<i,j>} t_{i,j} c^{\dagger}_{i\uparrow} c_{j\uparrow} + U \sum_{i} n_{i\uparrow}(n_{i\downarrow} - \frac{1}{2}) \quad (11)$$

$$\mathbf{H}^{\downarrow}_{D} = \sum_{i} V_i(t) c^{\dagger}_{i\downarrow} c_{i\downarrow} + \sum_{<i,j>} t_{i,j} c^{\dagger}_{i\downarrow} c_{j\downarrow} + U \sum_{i} n_{i\downarrow}(n_{i\uparrow} - \frac{1}{2}) \quad (12)$$

where $\mathbf{H}^{\uparrow}_{D}$ involves $n_{i\downarrow}$ and $\mathbf{H}^{\downarrow}_{D}$ involves $n_{i\uparrow}$. So these two Hamiltonians and the two sets of HEOM equations are coupled to each other by the Hubbard terms above.

## III. STEADY MAGNETIC PHASE TRANSITIONS

### A. Transitions in different device-lead coupling energies

We firstly investigate the density of states for the uniform QDA system in different coupling strength between the system and the two leads. The interacting system consists of odd number of QDs. From the previous works of QDA[18], the hopping integral energy and the Hubbard U energy are chosen as -27 meV and 20 meV respectively.

We see that in the weak coupling ($t_{couple}$ = -5.0 meV) case the system has the AF typed spin polarization (see inset of Fig. 1(a)). This can be further explained that in the DOS plot (Fig.1(a)). We see that below the Fermi energy (E=0), there are 5 spin-up DOS peaks and 6 spin-down DOS peaks. By integrating these peaks to the Fermi level, we find the number of the spin-up electron is 5.06 while the number of the spin-down electron is 5.94. Thus this result is consistent with the electron density distribution with a larger spin-down electron in the AF configuration.

While in the strong coupling case ($t_{couple}$ = -18 meV), the system has the same spin-up and spin-down electron density (0.5) and the same DOS curves for two spins. which means the QDA has no magnetism (The spin-up and spin-down electron number is 5.5 in this case). The reason for this magnetic to non-magnetic transition is that in the strong coupling case, some spin-up electron hops from the leads to the QDA device, to balance the majority of the spin-down electron. We find only when the coupling strength exceeds some critical values, can this hopping process occur. The detailed analysis is given later.

If the hopping energy is uniform among all QDs, this spontaneous spin polarization only exists in the odd-number-QD system. It is consistent with the Lieb's theorem and other's work about the carbon chain system [8, 11]. However, we find that if the hopping energy is not uniform, for example, the QD array with two alternative hopping values (such as the single-double-bond in the polyacetylene), for the even-number QDA, there may exist such spin polarization. Figure 1(c) and (d) shows the results for a QDA with N=10 and two alternative hopping energies: -27 meV and -20 meV. The Hubbard parameter is still 20 meV and the temperature is 3K. We see that in the weak coupling case ($t_{couple}$ = -5.0 meV) there exists AF-type spin polarization (inset of Fig. 1(c)) while the DOS curves of two spin cases coincide. Since this is even number system, the spin up electron distribution is symmetric to the spin-down electron distribution and their total numbers are the same. Similarly, in the strong coupling case ($t_{couple}$ = -18 meV) the spin-up and spin-down electron has the same uniform distribution (0.5) and there is no spin polarization for the QDA system (Fig. 1(d)).

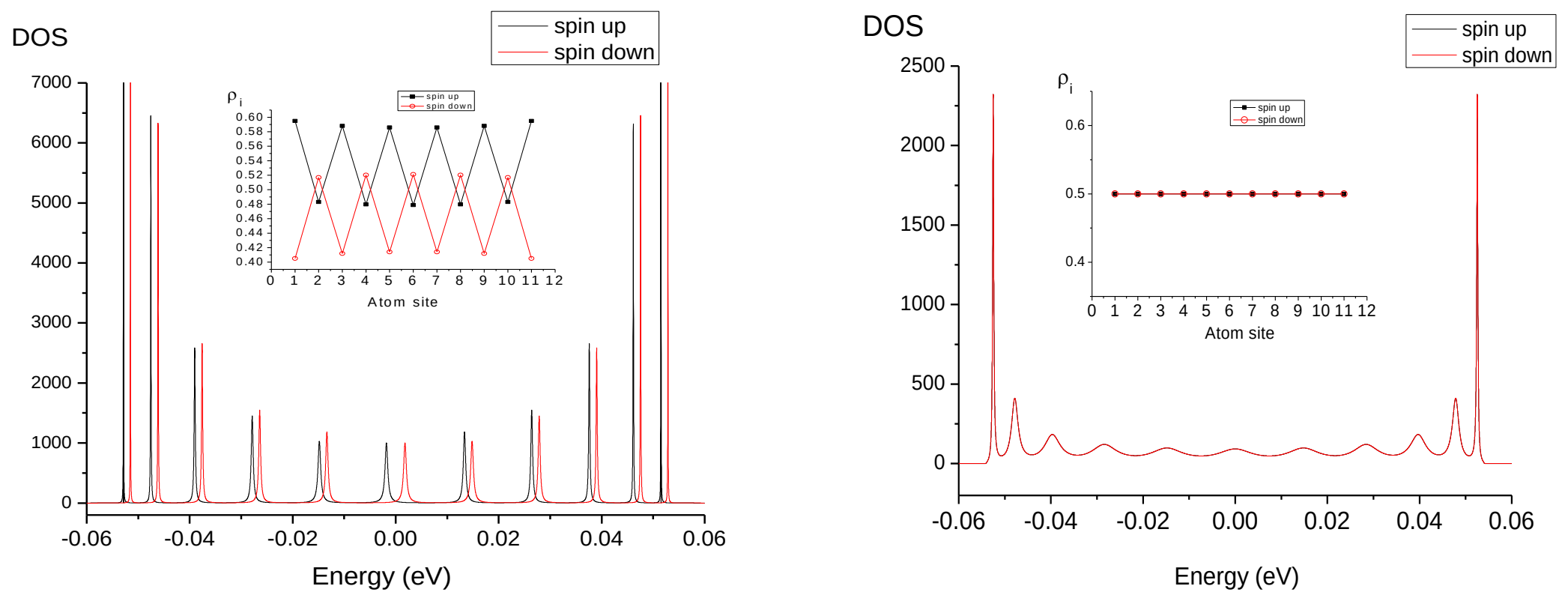

(a) (b)

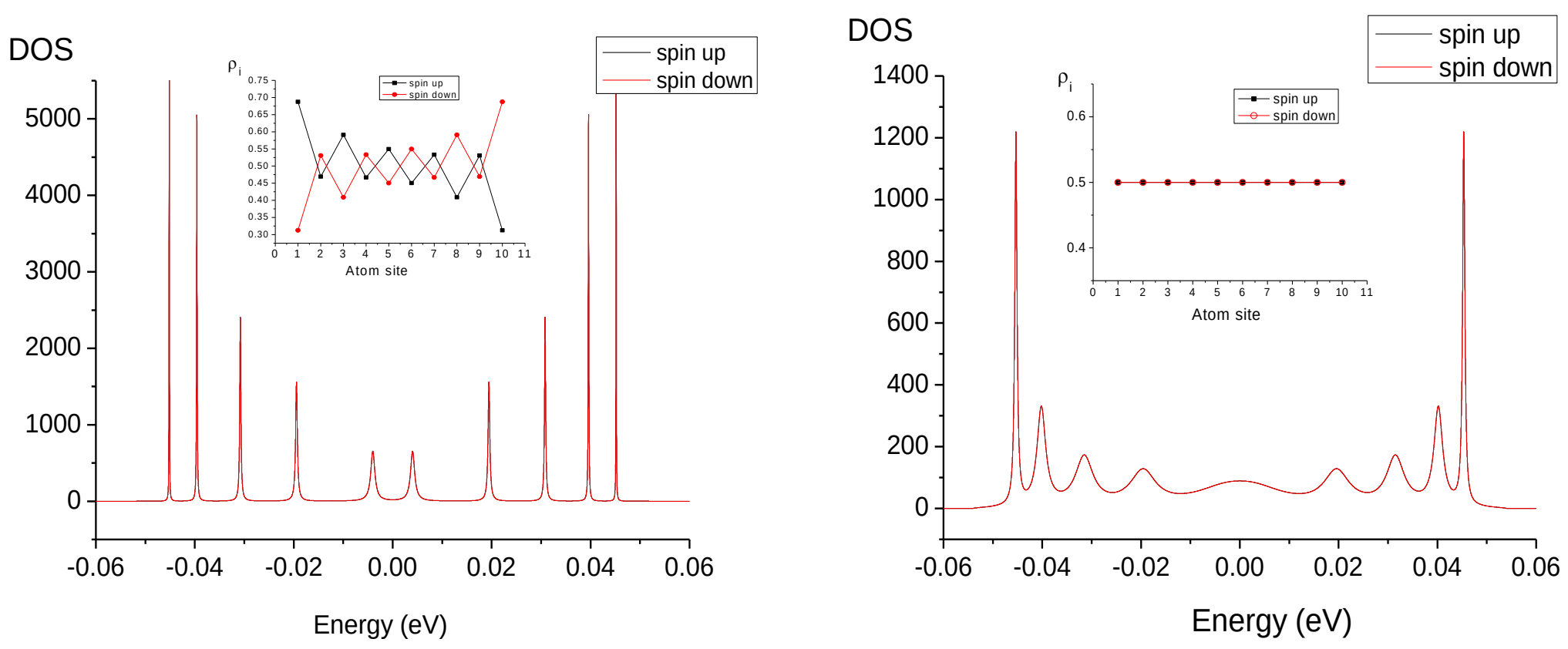

(c) (d)

**Fig. 1** DOS dependence on the energy and the electron distributions of the uniform-hopping QDA-lead system ((a) and (b), N=11, t=-27 meV) and the staggered hopping QDA-lead system ((c) and (d) N=10, $t_1$=-27 meV, $t_2$=-20 meV). (a) and (c) are for the weak coupling case ($t_{couple}$ = -5.0 meV) and (b) and (d) are for the strong coupling case ($t_{couple}$ = -18 meV). The corresponding electron distributions are in the inset of the figures.

Then we begin to investigate the phase transition of these QDA-lead systems with the change of coupling energy. From Fig. 2 (a) we see that when the coupling energy is larger than some critical value, the spin polarization gradually decreases to zero and the magnetism of the QDA-lead system (N=11) disappears. This is a type of 1$^{st}$ order phase transition due to the smooth change of spin polarization. It is also noticed that in a higher temperature, the critical coupling energy is lower. This is reasonable since in a high temperature the thermal fluctuation is larger, which makes the un-paired spin electron easier to hop from the leads to the device. We also observe the similar phenomena for the even-number QDA system with the staggered hopping energies (N=10). For this system the temperature dependence on the phase transition point is much weaker. In this QDA the spin-up and spin-down electron has the same number. The device-lead coupling only helps the electrons to hop more easily from one site to the neighboring sites, which eliminate the spin polarizations in each site. We believe in this case the thermal fluctuation plays less roles in the magnetic phase transition.

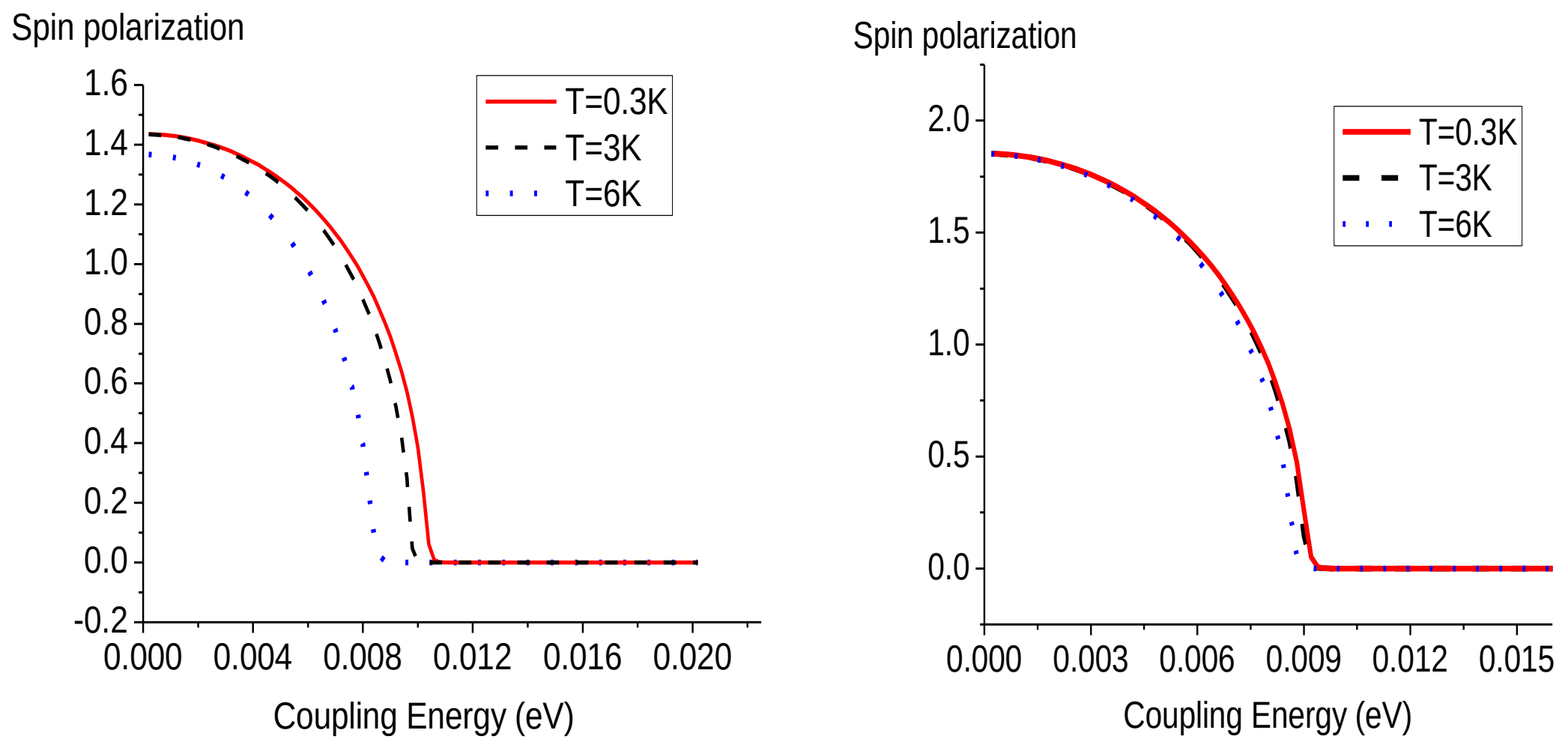

(a) (b)

**Fig. 2** The spin polarization dependence on the coupling strength for the QDA-lead system in the uniform hopping case ((a), N=11) and staggered hopping case ((b), N=10). The three curves in each subfigure correspond to different temperatures (0.3K, 3K and 6K).

### B. Verification by the exact diagonalization theory

We use the exact diagonalization technique to study the QDA system. The routine process of ED can be found in the literatures [18, 23]. We consider an odd-number QDA system (N=5) with the uniform hopping energy. With the parameters given below, we may calculation for the eigenvalues and electron density distribution for a state ($N_1$,$N_2$), which means in this state there are $N_1$ spin-up electrons and $N_2$ spin-down electrons.

Since N is odd, the numbers of spin-up and spin-down electron are not equal. Thus we have a spin polarization in this system. For the state (3,2) (or (2,3)) we do the average of all the eigenstates with the statistical weight factor $e^{-E_k/(k_B T)}/Z$, where Z is the partition function. In this (3,2) state, the electron distribution with spins are shown in Fig. 3(a). We see there exist AF-type spin polarization, which is very similar to the mean-field calculation by the NEGF theory with Hubbard model. For the symmetric state (2,3), the electron density distribution is the same, except the exchange of spin up and down components.

For the system embedded in the environment, the electron may jump out of (in) the system to (from) the environment (such as electrodes). The particle number is not conservative, but the chemical potential is conservative. People often use the grand ensemble (GE) statistics to describe the physical properties of the system [13, 18]. In GE, the statistical weights become $e^{-(E_k - n\mu)/(k_B T)} / Z$, where $\mu$ is the chemical potential with *n* particles. We here also gives the ground eigen-energies of different electrons in this QDA system, as shown in Fig. 3(b). In this figure we see that the state (3,2) and (2.3) has the same lowest total energy. As the chemical potential increases, we see in Fig. 3(c) that the average particle number also increases.

However, we find this ensemble average is not suitable for the magnetization of QDA here. If we use the ensemble average, that means we give each pair of spin complementary states (such as (3,2) and (2.3)) the same statistical weight. So after the ensemble average, their net spin polarization will be cancelled and there is no magnetic state. We realized that in the magnetic phase, the symmetry is broken: those symmetric spin complementation states do not simultaneously exist; or at least they do not have the same statistical weight. That has been validated by the free energy barrier between the two spin complementation states (Fig. 3(d)). In Fig. 3(d) we smoothly changes from state (3,2) to state (2,2) (by jumping off one spin-up electron to the leads), then to state (2,3) (by injecting one spin-down electron from the leads). In this intermedium process, we made a linear combination of these states and calculate its total energy. We see there indeed exist an energy barrier between the two symmetric states: (2,3) and (3,2).

From the NEGF results, we believe that when the lead-device coupling is strong enough, the state transfer between (3,2) and (3,2) becomes very frequently and then the system will lost the spin polarization.

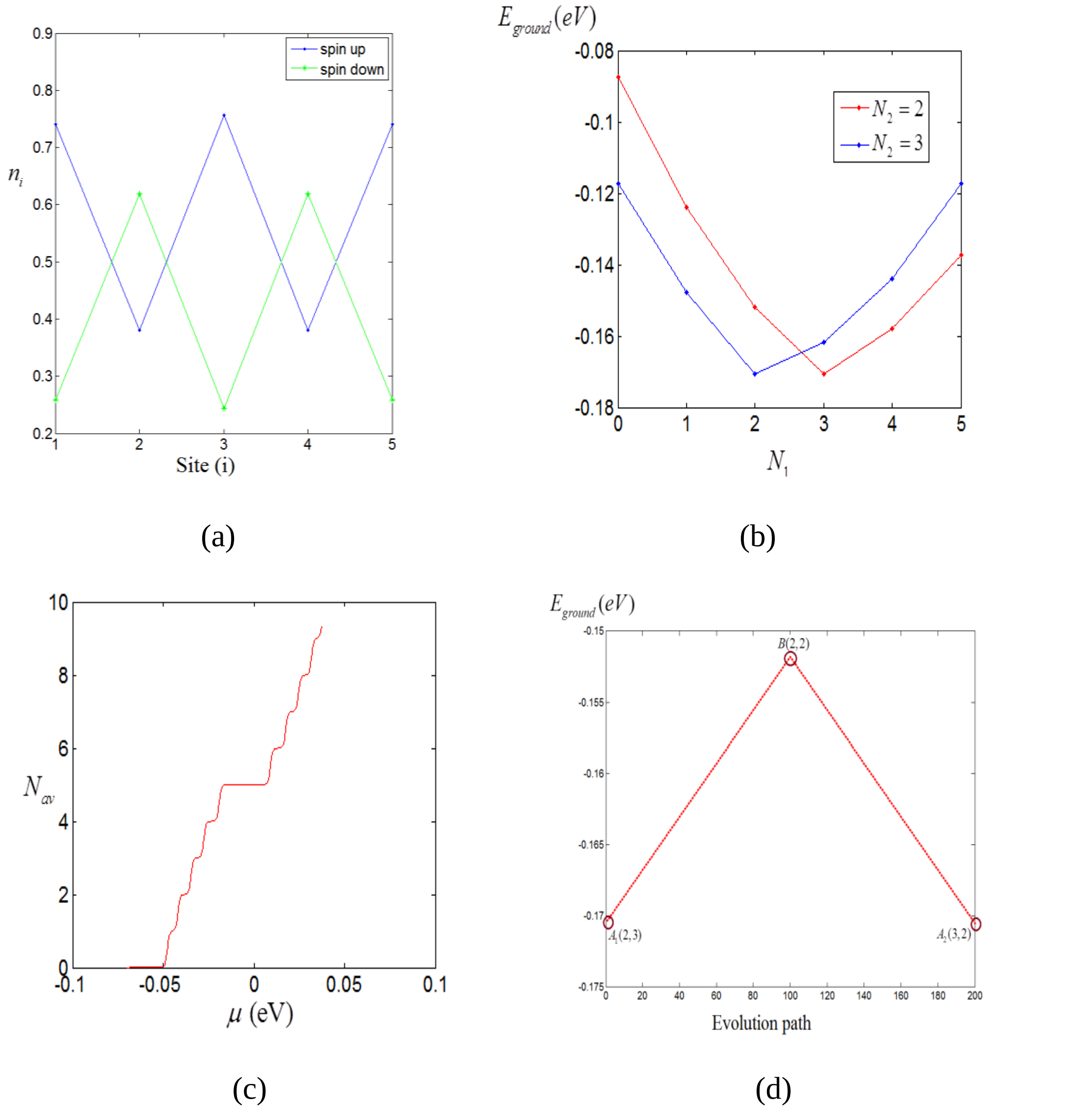

**Fig. 3** ED calculation results for the QDA system with N=5. The uniform hopping energy *t*=-10 meV; the Hubbard energy U=50 meV; the on-site energy $\varepsilon_0$=-30 meV. (a) The electron density distribution of state (3,2). The results are obtained from the average of all the eigen-states. (b) The average energy dependence on the number of spin-up electron in a canonical ensemble (N=5); (c) The average electron number of QDA with different chemical potentials. (d) The total energy change in a state evolution path: (2,3)—(2,2)—(3,2). In the intermedium process, the mixed state is obtained by a linear combination of these three states.

## IV. DYNAMIC MAGNEITC PHASE TRANISTIONS

### A. HEOM calculations

In this part we give the dynamic calculation for the open QDA system. The system still has three parts: left lead, central QDA and right lead. Here we use a larger hopping (uniform) and Hubbard energy: t=-2.7 eV, U=2.0 eV. The lead-device coupling energy is set as -0.8 eV. The detailed HEOM calculation method has been introduced in Sec. IIC and our previous works [19, 20]. For the Lorentzian and Pade expansion, we use $N_d$=4 Lorentzians and $N_p$=20 Pade points. We choose a QDA with N=11 and the steady state is obtained from the SC calculations. The electron distribution is very similar to Fig. 1(a). The steady state for the other auxiliary density matrices of two spins are obtained from the residue calculation as shown in our previous paper[20]. Then a symmetric bias with a stepwise temporal profile is applied on the two leads and we assume a linear potential drop between the two leads.

From Fig. 4 (a) we see that in the beginning there are huge current (and electron density) fluctuations due to the sudden application of bias voltage. And there are also transient spin polarized currents in the two leads, which come from the magnetism of the central QDA. At about 20 fs later, the spin-up and spin-down currents in the two leads tends to be the same and the net spin currents disappear. We also observe that when the system approaches to a steady state after about 100 fs, the electron distributions shows a non-magnetic state, as shown in Fig. 4 (b).

However, we find that if a small bias voltage is applied on this odd-numbered QDA system, the net spin current may sustain a much longer time (at about 60 fs, see Fig. 5(a)). The final electrons distribution in Fig. 5(b) shows that the system's magnetic property can survive after such small bias potential.

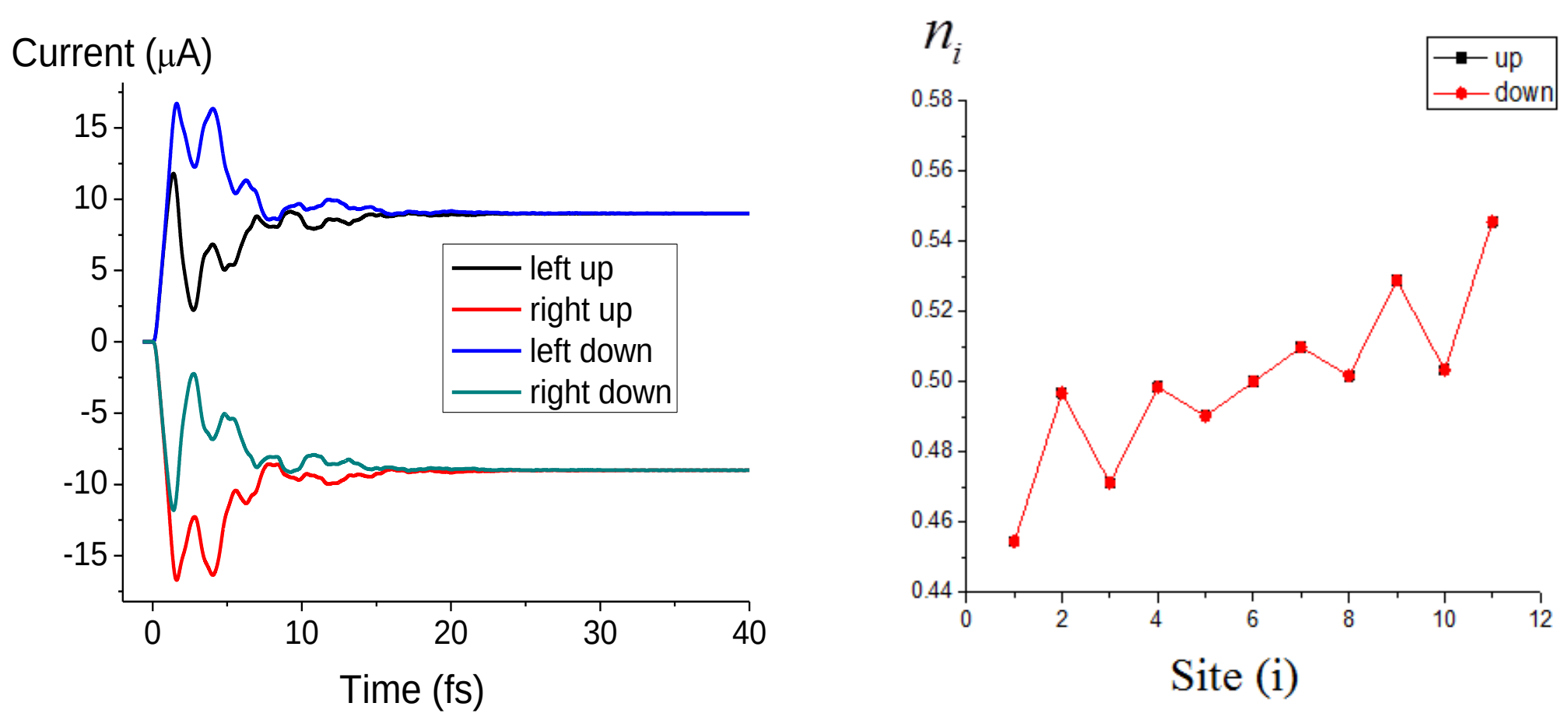

(a) (b)

**Fig.4** The dynamic magnetic phase transition process of an open QDA-lead system with a bias voltage of 1.0 V (N=11, hopping energy t=-2.7 eV, U=2.0 eV, weak coupling energy $t_{couple}$ = -0.80 eV). (a) Spin currents evolution of the two leads after the bias is applied; (b) Final stable electron distribution after about 100 fs.

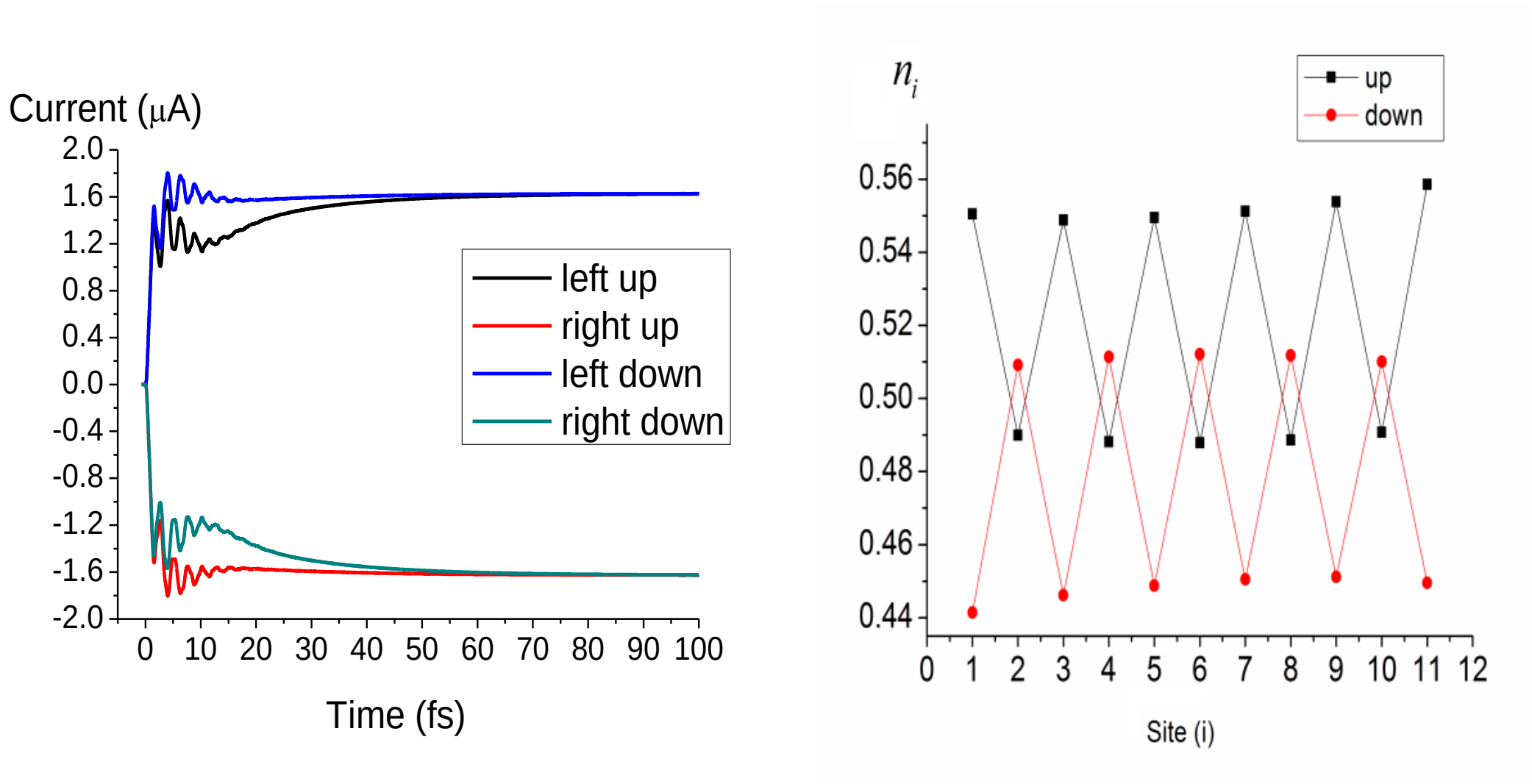

(a) (b)

**Fig. 5** The dynamic magnetic phase transition process of an open QDA-lead system with a bias voltage of 0.1 V (N=11, hopping energy t=-2.7 eV, U=2.0 eV, weak coupling energy $t_{couple}$ = -0.80 eV). (a) Spin currents evolution of the two leads after the bias is applied; (b) Final stable electron distribution after about 100 fs.

### B. Magnetic phase transitions in bias potentials

Then we use the NEGF theory to calculate the steady electron density distribution for the QDA (N=11) with a uniform hopping integral and weak coupling constant ($t_{couple}$ = -0.80 eV). The residue theorem is used for the energy integral for the equilibrium part while the direct integral is utilized for the non-equilibrium part. We find that when the bias is 1.0 V or 0.1 V, the electron distributions are the same as shown in Fig. 4(b) and Fig. 5(b) respectively. This coincidence verifies our dynamic calculation for this QDA system.

We then use this NEGF formula to calculate the spin polarization with different bias voltages. The result is in the following Fig. 6. This figure shows another magnetic phase transition for the weak coupling QDA system. When the bias exceeds some critical value, the electrons on the leads may jump into the QDA to balance the spin polarized sites, or the electrons in the spin polarized sites will run out of the QDA to the leads. All these processes eliminate the magnetism of QDA.

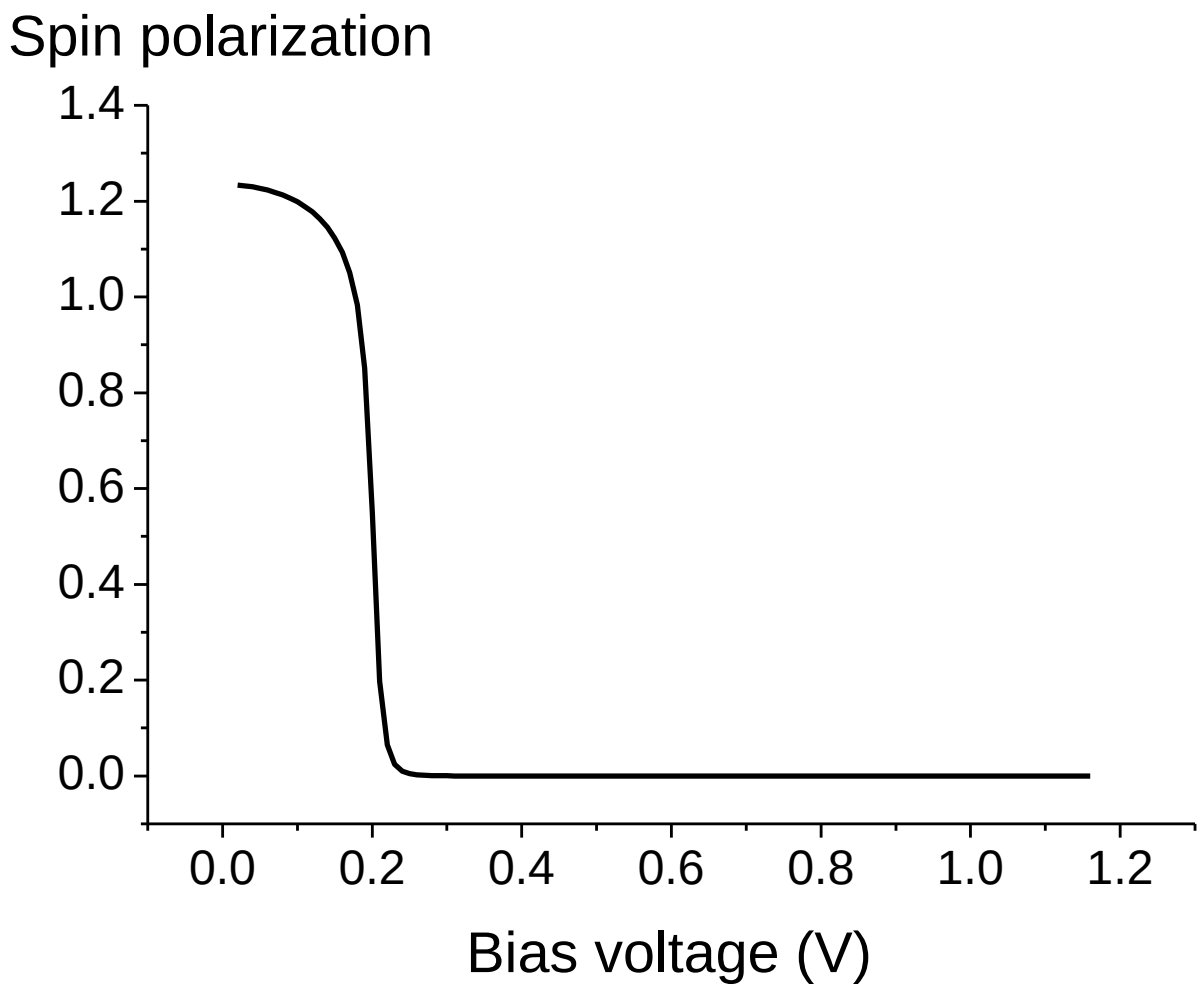

**Fig. 6** The spin polarization dependence on the bias voltage for the open QDA system with a uniform hopping integral and weak coupling (N=11, $t_{couple}$ = -0.80 eV).

## V. CONCLUSIONS

We use the Hubbard model with the NEGF theory to investigate a type of magnetic phase transition in the 1D interacting QDA coupled to non-interacting leads. When the coupling energy or the bias potential is beyond some critical value, the AF typed system turns into the nonmagnetic system. The AF order only exist in the odd-numbered QDA with a uniform hopping integral, or the even/odd-numbered QDA with staggered hopping integrals. ED calculation has verified this type of transition and the dynamic transition process has been studied by the HEOM method.

**ACKNOWLEDGEMENTS**

The authors thank Prof. Xuefeng Zhang in the department of physics, Chongqing University for his helpful discussions on the exact diagonalization. We thank Prof. Rui Wang for his kind help in the computer services. Financial support from the starting foundation of Chongqing University (Grants No. 0233001104429) and NSFC (11647307) are also gratefully acknowledged.